\documentclass[preprint,showpacs,showkeys,amsmath,amssymb,aps,doublespace]{revtex4}
\usepackage[dvips]{graphicx}
\usepackage{setspace}
\usepackage{amsmath}
\usepackage{amsfonts}
\usepackage{amssymb,color}

\usepackage{graphicx,color}
\usepackage{ifpdf}
\usepackage[latin1]{inputenc}

\begin{document}

\title{Inconsistencies in steady state thermodynamics}

\author{
Ronald Dickman\footnote{email: dickman@fisica.ufmg.br} and
Ricardo Motai\footnote{email: ricardotmotai@gmail.com}
}
\address{
Departamento de F\'{\i}sica and
National Institute of Science and Technology for Complex Systems,\\
ICEx, Universidade Federal de Minas Gerais, \\
C. P. 702, 30123-970 Belo Horizonte, Minas Gerais - Brazil
}

\date{\today}

\begin{abstract}

We address the issue of extending thermodynamics to nonequilibrium steady states.
Using driven stochastic lattice gases, we ask whether consistent definitions of
an effective chemical potential $\mu$, and an effective temperature $T_e$, are
possible.  $\mu$ and $T_e$ are determined via coexistence, i.e., zero flux of particles
and energy between the driven system and a reservoir.
In the lattice gas with nearest-neighbor exclusion, temperature is not relevant, and
we find that the effective chemical potential, a function of density and
drive strength, satisfies the zeroth law, and correctly predicts the densities of
coexisting systems.  In the Katz-Lebowitz-Spohn driven lattice gas both $\mu$ and
$T_e$ need to be defined.  We show analytically that in this case the zeroth law is
violated for Metropolis exchange rates, and determine the size of the violations
numerically.  The zeroth law appears to be violated for generic exchange rates.
Remarkably, the system-reservoir coupling proposed by Sasa and Tasaki
[J. Stat. Phys. {\bf 125}, 125 (2006)],
is {\it free} of inconsistencies, and the zeroth law holds.
This is because the rate depends only on the state of the
donor system, and is independent of that of the acceptor.
\end{abstract}


\maketitle

\section{Introduction}

Among the many open questions in nonequilibrium physics, a central issue
is whether thermodynamics can be extended
to systems far from equilibrium
\cite{oono-paniconi,hatano,hayashi,bertin,tomedeoliveira,evans}.
By ``thermodynamics" we mean a macroscopic description
employing a small number of variables, capable of predicting the final
state of a system following removal of some constraint \cite{callen}.
One expects the set of variables needed to describe a nonequilibrium system to be somewhat
larger than required for equilibrium; it should not, however, involve
microscopic details.  Thus in near-equilibrium thermodynamics,
fluxes of mass, energy, and other conserved quantities enter
as relevant variables \cite{onsager,degroot}.
This approach functions well in the hydrodynamic or local-equilibrium regime;
here, by contrast, we are concerned with systems maintained far from equilibrium.

Faced with a vast range of nonequilibrium phenomena, it is natural to focus on the
nearest analog to equilibrium thermodynamic states, that is, on nonequilibrium
steady states (NESS), and to analyze the simplest possible examples exhibiting such states,
for example, the driven lattice gas \cite{KLS,zia,marro}
or the asymmetric exclusion process \cite{schutz}.
Thus Sasa and Tasaki \cite{sasa2006}, extending the ideas of \cite{oono-paniconi},
proposed a general scheme of steady state thermodynamics (SST), including definitions
of the chemical potential and pressure in NESS, and developed a theoretical analysis of
the driven lattice gas;
a numerical implementation in driven systems is discussed in \cite{hayashi}.
More recently, Pradhan et al. \cite{pradhan}
tested, via numerical simulation of driven lattice gases in contact,
the consistency of the scheme proposed in \cite{sasa2006}, focussing on the
validity of the zeroth law.  These authors find that the zeroth law holds to good approximation,
and suggest that observed deviations may be attributed to the nonuniformities induced by the
contact itself.

A central notion in SST is that of {\it coexistence}.  Consider two systems, each in a steady state,
and weakly coupled to one another, so that they may exchange particles and/or energy.
We say that the systems coexist when the net flux of the quantity or quantities they may
exchange is zero.  If the two systems are in equilibrium states, then coexistence corresponds
to chemical and/or thermal equilibrium, marked by equality of their respective
chemical potentials and temperatures.  To construct a SST, we need
to define intensive parameters for NESS, such that the value
of the parameter associated with particle exchange (an effective
chemical potential, $\mu$) is the same when two systems coexist with respect to such exchange,
and similarly, an effective temperature, $T_e$, if the systems coexist with respect to energy exchange.
The definition of intensive parameters for nonequilibrium systems (such as the zero-range process)
possessing an asymptotic factorization property has been discussed in considerable detail by
Bertin et al. \cite{bertin}.

In this work we consider coexistence in classical lattice-gas models.
To eliminate inhomogeneities that might otherwise arise, we allow all sites to
participate in particle exchange.  In driven systems,
the rate $p_r$ of exchange attempts is a relevant parameter; we consider the limit
$p_r \to 0$, or the {\it weak exchange} limit.  By contrast, {\it virtual exchange}
characterizes the fluxes that would occur, were the systems allowed to exchange particles
and/or energy, given their respective stationary states; nothing is actually transferred.
The difference between virtual exchange and the
weak exchange limit is that in the latter case, fluxes are measured in a new stationary state,
attained under exchange of particles and/or energy, albeit at a vanishing rate.
Of course, null currents under virtual exchange is a necessary condition for having
null currents (and so coexistence) under weak exchange \cite{note1}.

Consider a pair of systems A and B, maintained at the same temperature, and free to exchange particles.
When the net flux of particles between the systems
is zero, we assert that their chemical potentials $\mu_A$ and $\mu_B$ must be equal,
if such functions in fact exist.
If one of the systems is a particle reservoir of known chemical potential,
we can use the zero-flux condition to measure the chemical potential of the other.
With $\mu$ so defined, we stipulate two essential
properties it must satisfy, to be a valid chemical potential:

1) (Zeroth law) If pairs of systems (A,B) and (A,C) separately satisfy the zero-flux
condition, then, if B and C are allowed to exchange particles, the net flux should
also be zero.

2) If systems A and B, at the same temperature, and initially isolated, with $\mu_A \neq \mu_B$,
are permitted to exchange particles, then the ensuing flux
should reduce $|\mu_A - \mu_B|$, and should continue until the difference is null.
In other words, knowing the functions $\mu_A(\rho)$ and $\mu_B(\rho)$, of two systems
in isolation should allow us to predict the direction of particle transfer when they are placed in contact,
and their densities at coexistence.

In the case of athermal lattice gases, the above considerations lead to a dimensionless
effective potential $\mu^* \equiv \mu/k_BT$ equivalent to that defined by Sasa and Tasaki \cite{sasa2006}; we show
in Sec. III that this function satisfies the two criteria stated above.
For the lattice gas
with nearest-neighbor interactions (the Katz-Lebowitz-Spohn or KLS model),
one must decide how the rates of particle exchange between systems depend
on their respective changes in energy; even restricting attention to
rates satisfying detailed balance, there are many possible choices.
We show that for the familiar Metropolis rates, the simplest extension of the
approach used in athermal systems does not predict coexistence correctly.
This leads us to define an effective temperature as well as an effective
chemical potential for the driven lattice gas.
We shall nevertheless show that such a definition in general leads to violations
of the zeroth law.  Remarkably, the exchange scheme proposed by Sasa and Tasaki (ST)
is free of such inconsistencies.  Using this approach, it is possible to define
$\mu^*$ in a consistent manner, and to predict coexisting densities
in the KLS model.  The essential difference between ST rates and more familiar
expressions such as Metropolis rates is that in the former, the rate for transferring
a particle from system A to system B depends on the change in energy of A, but not
on that of B.

The balance of this paper is organized as follows.  In Sec. II we review the
kinetics of particle transfers, focussing on the properties of the transition rates.
Section III discusses the application to the NNE lattice gas.
In Sec. IV, a direct application of the method to the KLS model
(involving only the effective chemical potential), is shown to fail when
Metropolis exchange rates are used.  Coexistence with respect to particle
exchange is correctly predicted using Sasa-Tasaki rates.  In both cases,
however, there is a steady flow of energy from the driven to the nondriven system,
motivating the definition of an
effective temperature.  The consequences of this definition are
investigated analytically and numerically in Sec. V.
We close in Sec. VI with a
discussion of the implications of our results for steady state thermodynamics.

\section{Transition rates}

Since our definitions of intensive parameters on based on stationary averages of fluxes of
particles and energy between systems, it seems worthwhile to review briefly the kinetics
of particle transfers.  We consider classical stochastic lattice gases, that is, Markov processes
whose state space corresponds to possible configurations of particles on a lattice or
a pair of lattices, and
whose evolution is defined by transition rates between certain pairs of configurations.
We denote the transition rate from configuration ${\cal C}$ to ${\cal C}'$ by
$w({\cal C}'|{\cal C})$.

In this work we are concerned with transfers of particles between systems.  Let $A$ and $B$
be lattice gases characterized by intensive parameters $\mu_A$ and $\beta_A \equiv 1/k_B T_A$,
and similarly for $B$.  Let ${\cal C}_A$ be a configuration of $A$, with
energy $E_A$, and let ${\cal C}'_A$, with energy $E'_A = E_A + \Delta E_A$,
be a configuration of $A$ obtained by adding a single particle to ${\cal C}_A$.
Consider now the composite system $(A,B)$, with initial configuration ${\cal C} \equiv ({\cal C}_A, {\cal C}'_B)$,
and let ${\cal C}' \equiv ({\cal C}'_A, {\cal C}_B)$ be a configuration of the composite system
after transferring a particle from $B$ to $A$.
Although the systems of interest are not necessarily
in equilibrium, we need to include such systems in our description, and shall therefore
assume that the transition rates for particle transfers between systems satisfy detailed
balance. That is, if $\beta_A = \beta_B = \beta$,

\begin{equation}
\frac{w({\cal C}'|{\cal C})}{w({\cal C}|{\cal C}')} = \exp [\beta(\mu_A - \mu_B) - \beta(\Delta E_A - \Delta E_B)].
\label{detbal1}
\end{equation}

\noindent In case system $B$ is a particle reservoir with dimensionless chemical potential
$\mu^* \equiv \beta \mu$, we ignore
the associated reservoir configurations and simply write,

\begin{equation}
\frac{w({\cal C}'_A|{\cal C}_A)}{w({\cal C}_A|{\cal C}'_A)} = \exp [\mu^* -  \beta \Delta E_A ].
\label{detbal2}
\end{equation}

As is well known, there are many possible choices of transition rates consistent with detailed
balance.  In studies of equilibrium systems, all such choices are equivalent insofar as static properties
are concerned, and rates can be chosen on the basis of
calculational convenience or computational efficiency. Out of equilibrium, however, the choice of rates
can have very significant effects on stationary and time-dependent behavior.  This holds both for the
rates governing the {\it internal dynamics} of a system, and for particle exchanges between systems.
In principle, one would like to have a SST that is valid for any choice of rates.  But we shall see that
the consistency condition discussed in the Introduction implies a strong constraint
on the rates for particle transfers between systems.

Perhaps the most familiar rate function satisfying detailed balance is that associated with
{\it Metropolis rates}, for which

\begin{equation}
w_M({\cal C}'|{\cal C}) = \epsilon \min \{1, \exp [\beta(\mu_A - \mu_B) - \beta(\Delta E_A - \Delta E_B)]\},
\label{metro}
\end{equation}

\noindent where $\epsilon$ is an arbitrary, constant rate.
Another function widely studied in the context of driven lattice gases the so-called
{\it mechanism-B rate}:

\begin{equation}
w_B({\cal C}'|{\cal C}) = \frac{\epsilon}{1+  \exp [\beta(\mu_A - \mu_B) - \beta(\Delta E_A - \Delta E_B)] }.
\label{mecB}
\end{equation}
\vspace{1em}

A rather less studied rate (in simulations) is that introduced in \cite{sasa2006}, which we shall
call {\it Sasa-Tasaki} or ST rates:

\begin{equation}
w_{ST}({\cal C}'|{\cal C}) = \epsilon  \exp [\beta (\mu_A - \Delta E_A)] .
\label{ST}
\end{equation}

\noindent The rate for the reverse transition is obtained by letting $A \to B$.  Thus the rate for
particle transfer from $A$ to $B$ depends exclusively on parameters associated with $A$, and vice-versa.
We shall refer to this important condition as the {\it ST property}.  The class of rates satisfying detailed
balance and the ST property is unique to within a multiplicative factor, $\epsilon$.
The ST expression can be seen as resulting from a very high energy barrier $W$ between $A$ and $B$; the
particle first makes a transition from $A$ to the barrier, and from there to $B$.  In this
case the prefactor $\epsilon \propto e^{-\beta W}$; the limit $W \to \infty$ corresponds to the
weak exchange limit  \cite{kramers}.

\section{Lattice gas with nearest-neighbor exclusion}

To begin we examine the lattice gas with nearest-neighbor
exclusion (NNE).  In this model the pair interaction is infinite if the distance between particles
is $\leq 1$, and is zero otherwise.  Since there is no characteristic energy scale,
the relation between the density and the dimensionless chemical potential $\mu^* = \mu/k_BT$ is independent of
temperature.  Such models are termed {\it athermal}.
The NNE lattice gas has been studied extensively as a discrete-space version
of the hard-sphere fluid \cite{runnels,gaunt,ree,fernandes}, and is known to
exhibit a continuous (Isinglike) phase transition to sublattice
ordering at a density of $\rho_c \simeq 0.36774$ \cite{guo}.

Consider a NNE system on a lattice of $L^d$ sites, with a fixed number of particles,
and with a stochastic dynamics (e.g., particle hopping) obeying detailed balance.
This simply means that if ${\cal C}$ and ${\cal C}'$ are two valid configurations,
then the transition probabilities $P({\cal C}'|{\cal C})$ and $P({\cal C}|{\cal C}')$
must be equal.
Any particle displacement satisfying the NNE condition is accepted.
Suppose the system is also allowed to exchange particles with a reservoir at dimensionless
chemical potential $\mu^*$. Exchanges may be
implemented as follows.  A site is selected at random, and if it is occupied, the particle is removed
with a certain probability $p_R$. If the chosen site is {\it open},
i.e., vacant and with all nearest neighbors also vacant, then a particle is inserted with probability $p_I$.
If ${\cal C}'$ is a configuration obtained by adding a single
particle to configuration ${\cal C}$, then at equilibrium the respective probabilities satisfy
$P({\cal C}')/P({\cal C}) = e^{\mu*}$, which implies $p_I/p_R = e^{\mu*}$.

For a system with $n$ particles, the probability of gaining a particle
in an exchange is,
\begin{equation}
P(n \to n+1) = \sum_{\cal C} \frac{n_{op}({\cal C})}{L^d} p_I P({\cal C}) ,
\label{pgain}
\end{equation}
\noindent where the sum is over all $n$-particle configurations, and $n_{op}({\cal C})$ is
the number of open sites in configuration ${\cal C}$.
Similarly, the probability of losing a particle is
\begin{equation}
P(n \to n-1) = \sum_{\cal C} \frac{n}{L^d} p_R P({\cal C}).
\label{ploss}
\end{equation}
\noindent
The coexistence condition $P(n \to n+1)=P(n \to n-1)$
implies
$\mu^* = \ln (\rho/\rho_{op})$,
where $\rho_{op}$ is the average density of open sites over configurations with $n$ particles.

The NNE lattice gas exhibits
nonequilibrium steady states under a drive \cite{drnne}.
In the simple case of single-particle nearest-neighbor hopping on the square lattice, we
parameterize the drive such that hopping attempts in the
$+x$ direction occur with probability $p/2$, and those in the opposite direction with
probability $(1\!-\!p)/2$;
the attempt probabilities for hopping in the $\pm y$ directions are 1/4 as in equilibrium.
Then $p=1/2$ corresponds to the unbiased (equilibrium) case, and $p=1$ to maximum drive.
The model exhibits a line of phase transitions in the $\rho - p$ plane, with
$\rho_c \simeq 0.263$ for $p=1$.

We define $\mu^*$ in the {\it driven} NNE lattice gas (and in any
driven athermal system) via the same relation: $\mu^* \equiv \ln (\rho/\rho_{op})$.
The reason is that the reservoir
must have this value of $\mu^*$ for the net particle flux between it
and the system (driven or not) to be zero.
Note that our definition of $\mu^*$ is equivalent to
the general definition proposed by Sasa and Tasaki \cite{sasa2006}.

This definition of $\mu^*$ is compatible with the zeroth law for all values of the
drive.  To see this, consider two systems,
${\cal S}$ and ${\cal S}'$, characterized by densities $\rho$ and $\rho'$,
and drive parameters $p$ and $p'$, and which satisfy the zero-current condition with respect to
the same particle reservoir, so that $(\rho/\rho_{op}) = (\rho'/\rho'_{op})$.  Under
virtual exchange between ${\cal S}$ and ${\cal S}'$, the net particle
current is
$\langle \Delta n \rangle_{{\cal S}{\cal S}'} = \rho' \rho_{op} - \rho \rho'_{op}$,
which is identically zero under the hypothesis of a common value of $\mu^*$.

To determine the effective chemical potential $\mu^* (\rho,p)$ of the driven NNE lattice gas,
we simulate the model on square lattices of size $L \times L$ with $L=40$, 80, and 100,
with periodic boundaries.
(Restricting the analysis to the disordered phase, $\rho < \rho_c$, finite-size effects are
weak.)  We use a set of 5-10 independent realizations, each consisting of $2 \times 10^5$ lattice
updates (LU) for relaxation, followed by $10^7$ LU for calculating stationary averages. (A LU
corresponds to one attempted move per site.)

A second series of studies probes coexistence, by simulating a pair of systems,
A and B, of the same size. At each event, particle exchange is selected with probability $p_r$.
We use a series of small $p_r$ values (typically, 0.001-0.006)
to allow extrapolation to $p_r =0$.
Exchange is allowed between any pair of sites, $(i,j)_A$ and $(i',j')_B$ in the two lattices.
In an exchange move, a pair of sites is chosen at random.  Then
if $(i,j,)_A$ is occupied
and $(i',j')_B$ is open, or vice-versa,
a particle is exchanged between the systems.

\begin{figure}[!htb]
\includegraphics[clip,angle=0,width=0.8\hsize]{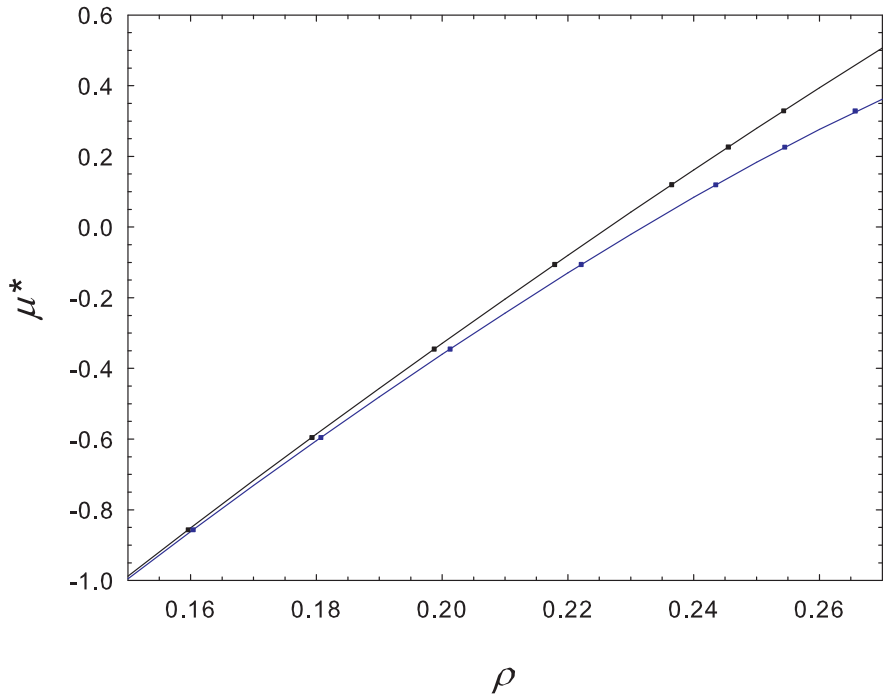}
\vspace{-3cm}

\caption{NNE lattice gas: simulation results for $\mu^*$ in equilibrium (black curve)
and under maximum drive (blue curve), system size $L=80$.  Pairs of points sharing the
same value of $\mu^*$ represent coexisting densities in the equilibrium and driven
systems under weak exchange.  Uncertainties are smaller than line thickness and symbol size.}
\label{drcx80}
\end{figure}

Simulation results for $\mu^* (\rho,p)$ in isolated systems are shown in Fig.~\ref{drcx80} (smooth curves).
(The system size is $L=80$; similar results are obtained for other sizes).
We see that for a given density, $\mu^*$ under maximum drive is smaller than in equilibrium,
and that the difference increases with density.
This can be understood as the result of a bunching effect: particles tend to pile up along the drive,
leaving larger vacant regions and hence a higher value of $\rho_{op}$, which in turn
reduces $\mu^*$.
The points in Fig.~\ref{drcx80}
show pairs of coexisting stationary densities (extrapolated to $p_r=0$), between a system with $p=1$
and one in equilibrium.
Each pair of coexisting systems possesses a single value of $\mu^*$ (to within a statistical
uncertainty of $10^{-4}$ or less), showing that equality of this quantity can be used to predict the
coexisting densities.  The points, moreover, fall along the stationary $\mu^* (\rho,p)$ curves
of the corresponding isolated systems, confirming that, under weak exchange, the systems are governed
by the same relation between the intensive variables $\rho$ and $\mu^*$ as they are in isolation.

\begin{figure}[!htb]
\includegraphics[clip,angle=0,width=0.8\hsize]{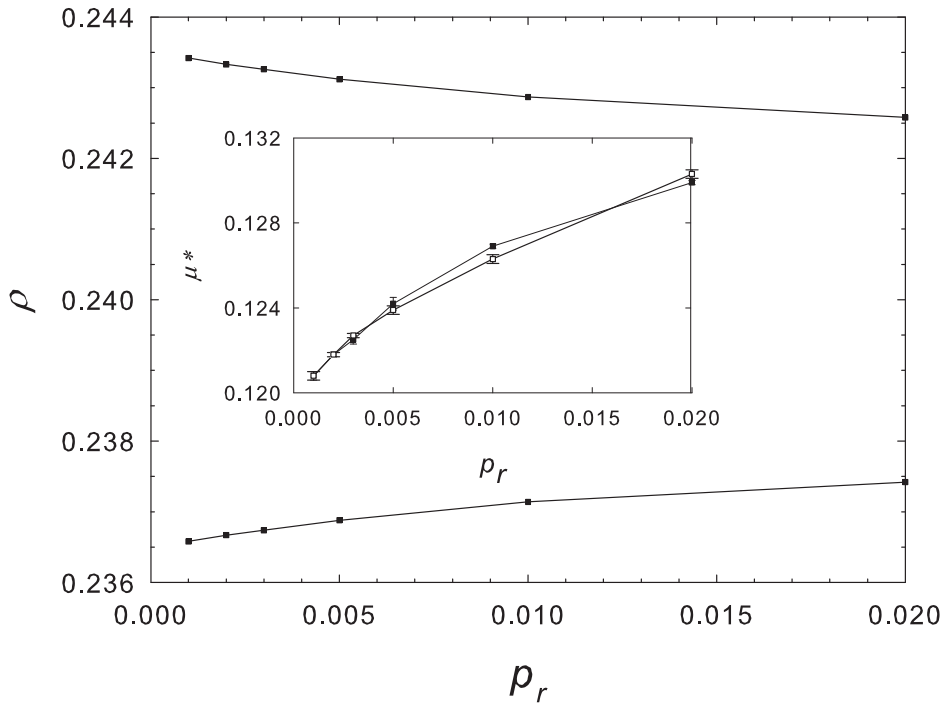}
\vspace{-3cm}

\caption{NNE lattice gas: simulation results for coexisting densities in a nondriven system (lower)
and under maximum drive (upper) as a function of exchange rate $p_r$ (system size $L=80$, mean density
$\overline{\rho} = 0.24$).  Inset:
chemical potentials $\mu^*$ at coexistence (for the same parameters)
in the nondriven (squares) and driven ($\times$) systems.  Error bars are smaller than symbols.}
\label{drmu824a}
\end{figure}

As noted in the Introduction, stationary properties at coexistence depend on the rate $p_r$ of
exchange attempts.
This is illustrated in Fig.~\ref{drmu824a}: both the coexisting densities
and chemical potentials vary systematically with $p_r$, converging to well defined limits
as $p_r \to 0$; the limiting chemical potential values agree to within uncertainty.
[For the parameters of Fig.~\ref{drmu824a}, the limiting values are $\mu^* = 0.11957(6)$
and 0.11959(10) in the nondriven and driven systems, respectively.]
This dependence arises because varying $p_r$ changes the fraction of moves
in which the drive acts, effectively varying its strength.
Increasing $p_r$ reduces the effect of the drive, so that
the coexisting densities approach one another, and $\mu^*$ increases, approaching its equilibrium
value.  (As a technical point we note that although very small values of $p_r$ would in principle
be desirable in approaching the limit, such values also imply slow relaxation and large uncertainties
in simulations, so that in practice it is better to perform studies at a series of reasonably
small $p_r$ values and extrapolate to $p_r = 0$.)

The zeroth law is verified in the following manner.  We first simulate a pair of systems, one (A) in equilibrium,
the other (B) with maximum drive, under weak exchange.  For the parameters used ($L=100$, mean density
$\overline{\rho} = 0.26$), we obtain the coexisting densities
$\rho_A = 0.2542(1)$ and $\rho_B = 0.2658(1)$.  The measured
chemical potential is $\mu^* = 0.3278(8)$ in both systems.
We next examine a system (C)
with drive $p=0.6$, and find that it coexists with system A (at density $\rho_A = 0.2542$, as above)
when $\rho_C = 0.2548(1)$; the corresponding chemical potential values agree to within uncertainty
with that obtained under coexistence of systems A and B.  The zeroth law then implies that
systems B and C should also coexist at the previously obtained densities.  This is verified by
simulating these systems under weak exchange, starting each with density $\overline{\rho} = 0.2603$.
In the stationary state, the coexisting densities are $\rho_C = 0.2548(1)$ and $\rho_B = 0.2658(1)$,
just as expected.  We have repeated this analysis using other densities and drive values, confirming
in each case the validity of the zeroth law.

Given that $\mu^*$ is an increasing function of density, for any value of the drive,
our analysis shows that if NNE models with different values of $\mu^*$ are permitted
to exchange particles, the ensuing particle flux will tend to equalize the chemical potentials.
Since our definition of the effective chemical
potential is based on the particle insertion probability, which is well defined even under a drive,
we expect that steady state thermodynamics can be defined in a consistent manner for
athermal systems in general, at least insofar as configuration space is concerned.

\subsection{NNE lattice gas with first- and second-neighbor hopping}

\begin{figure}[!htb]
\includegraphics[clip,angle=0,width=0.8\hsize]{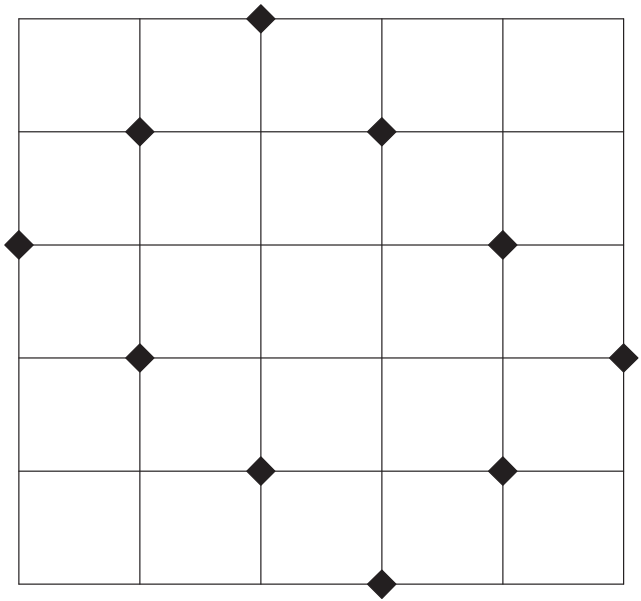}
\vspace{-3cm}

\caption{NNE lattice gas: example of a particle configuration satisfying the NNE condition,
but which admits no escape via nearest-neighbor hopping.}
\label{trapped}
\end{figure}

Although our numerical results are in good accord with theoretical expectations
for the driven NNE lattice gas with nearest-neighbor hopping, the model has
a defect that might be expected to complicate theoretical analyses: it is
{\it nonergodic}.  Specifically, there are valid configurations which admit
no escape via hopping to nearest-neighbor sites; an example of such a
``trapped" configuration is shown in Fig.~\ref{trapped}.  (Note that trapped
configurations exist already in equilibrium; for $p=1$ the class of trapped configurations
is considerably larger.)  Since (for $p<1$) a trapped configuration ${\cal C}^*$ is inaccessible
from any other configuration, the evolution cannot visit the full configuration space,
making the process nonergodic.  (For $p=1$ the ergodicity violation is stronger,
since, starting from a generic initial configuration, the process can become trapped
in an absorbing configuration \cite{drnne}.)

If we extend the set of transitions to include hopping to second as well as first neighbors,
the process appears to be ergodic for all values of $p$; we shall refer to this as the
NNE2 lattice gas.
The hopping rates are 1/8 in the directions perpendicular to the drive, $p/4$ for
displacements with a positive projection along the drive, and $(1\!-\!p)/4$
for those with a negative projection.
The phase diagram of driven NNE2
model was studied some years ago by Szolnoki and Szabo \cite{drnne2}, who showed that
there is a line of Ising-like phase transitions in the $\rho - p$ plane, with
$\rho_c \simeq 0.35$ for $p=1$.  As before, we restrict attention to the disordered
phase.

\begin{figure}[!htb]
\includegraphics[clip,angle=0,width=0.8\hsize]{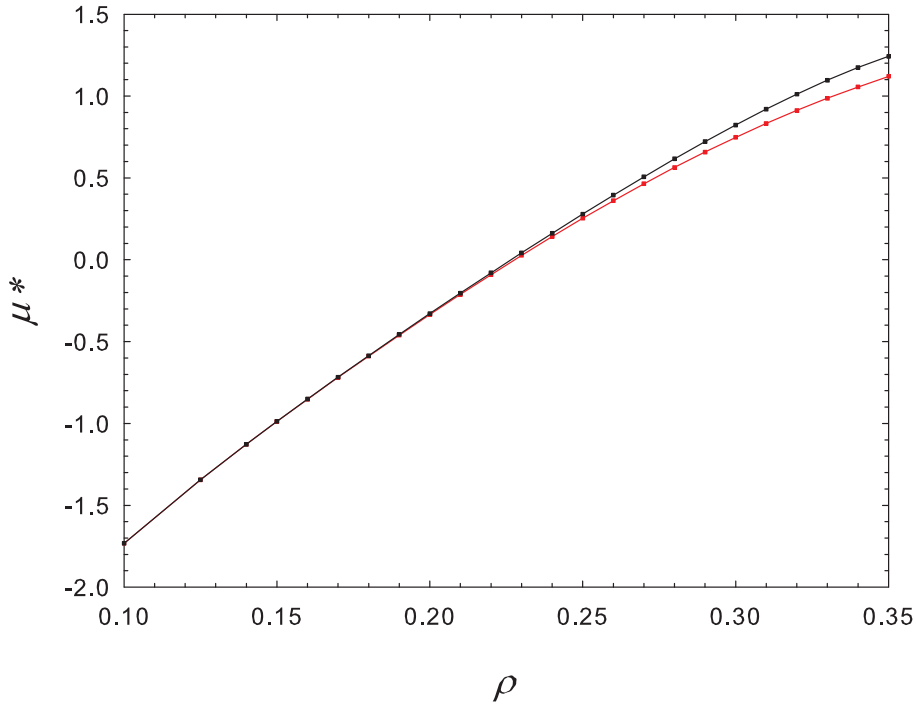}

\caption{NNE2 lattice gas: dimensionless chemical potential $\mu^*$ versus density for
$p=1/2$ (upper curve) and $p=1$ (lower curve).  System size $L=80$.}
\label{munn2}
\end{figure}

The behavior of $\mu^*$ in the driven NNE2 model is qualitatively similar to that of
the model with only nearest-neighbor hopping (the NNE model).  The dimensionless chemical potential for
$p=1$ is plotted versus density in Fig.~\ref{munn2}.  The difference between
$\mu^*(\rho,p=1)$ and $\mu^*(\rho,p=1/2)$ grows more slowly with density than in the NNE model,
as might be expected, since the tendency to block particle motion is reduced in the present case.

As in the NNE model, we verify that under weak exchange between systems with different values
of the drive, $p$ and $p'$, the coexisting densities $\rho$ and $\rho'$ are predicted by equating
the chemical potentials $\mu^*(\rho,p)$ and $\mu^*(\rho',p')$.  For example,
simulations of isolated systems (with system size $L=80$) in equilibrium and under maximum drive yield
a common value of $\mu^*$ for particle numbers 2040 and 2124 (densities 0.31875 and 0.33188), respectively.
Simulations of the two lattice system, using $p_r = 0.001$ - 0.005 yield (in the limit $p_r \to 0$)
the same chemical potential, $\mu^* = 1.0000(1)$, and coexisting densities 0.31876(1) and 0.22186(1).
Thus consistency of SST is verified in the NNE2 lattice gas, supporting our surmise that it holds for
athermal systems in general.

\section{Driven lattice gas with attractive interactions}

We turn our attention to the driven lattice gas
(KLS model) with attractive nearest-neighbor interactions.
The system evolves via a particle-conserving dynamics with a drive ${\bf D}= D{\bf i}$ favoring
particle displacements along the $+x$ direction and inhibiting those in the opposite sense.
The acceptance probability for a particle displacement $\Delta {\bf x}$ is
\begin{equation}
p_a = \min\{1, \exp[-\beta(\Delta E - {\bf D} \cdot \Delta {\bf x})]\}.
\label{padlg}
\end{equation}

\noindent In this case, we consider nearest-neighbor hopping, which is sufficient to ensure
ergodicity.

In this section we attempt to describe coexistence between KLS systems using
the approach employed for the NNE lattice gas.  That is, given a system ${\cal S}$ with
density $\rho_{\cal S}$, temperature $T$ and drive $D>0$, we determine
$\mu_{\cal S}^*$ via the zero-current condition.  We then examine the possibility of
coexistence between ${\cal S}$ and a nondriven (equilibrium) system ${\cal S}_0$ at temperature $T$,
whose density $\rho$ is adjusted to render its chemical potential $\mu_0^*$ equal to
$\mu_{\cal S}^*$.

Since particle exchanges with the reservoir also transfer energy, we must alter the
insertion and removal probabilities; a convenient choice satisfying detailed balance is:

\begin{equation}
p_{I} = \min [1, e^{\beta[\mu - (E_{new} - E_{cur})]} ],
\label{paI}
\end{equation}
\noindent and
\begin{equation}
p_{R} = \min [1, e^{\beta[-\mu - (E_{new} - E_{cur})]} ].
\label{paR}
\end{equation}
\noindent In the above expressions, $E_{cur}$ is the energy of the current configuration and $E_{new}$
that following the change, be it insertion or deletion.
The above expressions correspond to Metropolis rates, as discussed in Sec. II.

Given that the system is in configuration ${\cal C}$, the expected change
$\langle \Delta n \rangle_{\cal C}$ in the particle number $n$ in an attempted
insertion or removal is
$\langle \Delta n \rangle_{\cal C} = \mbox{Prob} [n \to n+1] - \mbox{Prob} [n \to n-1]$.  For the procedure
defined above,

\begin{equation}
\langle \Delta n \rangle_{\cal C} = \frac{1}{L^d} \sum_j  [(1-\sigma_j) p_{I;j} - \sigma_j p_{R;j}]
\label{deltan}
\end{equation}

\noindent where the sum is over sites and $\sigma_j = 1(0)$ if site $j$ is occupied (vacant).
The probabilities $p_{I;j}$ and $p_{R;j}$ are those of Eqs.~(\ref{paI}) and (\ref{paR}),
respectively, with $E_{new}$ the energy following insertion or removal at site $j$.
If $P({\cal C})$ is the probability
distribution on configuration space,
then the expected change in particle number per attempted transfer
(insertion or removal) is

\begin{equation}
\langle \Delta n \rangle_P = \sum_{\cal C} P({\cal C}) \langle \Delta n \rangle_{\cal C}.
\label{DeltanP}
\end{equation}

In the KLS model, the change in energy upon particle insertion (removal) at a site with $j$ occupied
neighbors is $\Delta E = -j$ ($\Delta E = +j$).  Thus the expected change in particle number per exchange event is

\begin{equation}
\langle \Delta n \rangle = \sum_{j=0}^q \left[ \rho_{\cal S}^+(-j) \min\{1,e^{\beta_e(\mu + j)}\} -
                             \rho_{\cal S}^-(j) \min\{1,e^{-\beta_e(\mu + j)}\} \right],
\end{equation}

\noindent where
$q$ is the lattice coordination number, and $\rho_{\cal S}^+(-j)$ [$\rho_{\cal S}^-(j)$]
is the density of vacant [occupied] sites
with exactly $j$ occupied nearest neighbors.  The densities are stationary averages over
the set of configurations with $n$ particles, as in the NNE case.
Setting $\langle \Delta n \rangle_P = 0$, we have an equation for $e^{\mu^*}$.

On the square lattice, the allowed
values of $\Delta E$ are -4, -3,..., and 0,
while for removal $\Delta E$ takes the values 0, 1,...,4.
Consider first $\mu \geq 0$.  In this case, $p_{I;\Delta E} = 1$ for all pertinent values of $\Delta E$, while
$p_{R;\Delta E = j} = e^{-\beta (\mu +j)}$ for $j=0,1,...,4$.
Then Eq. (\ref{DeltanP}) becomes:
\begin{eqnarray}
\langle \Delta n \rangle_P
= \{\!\!\! &(& \!\!\! 1 - \rho) - e^{-\mu^*} [\rho^-(0) + e^{-\beta} \rho^-(1)
\nonumber
\\
&+& e^{-2\beta} \rho^-(2) + e^{-3\beta} \rho^-(3) + e^{-4\beta} \rho^-(4)
]\},
\label{deltanmp}
\end{eqnarray}

\noindent where we used the fact that $\sum_{j=0}^4 \rho^+(j)$ is just the density of vacant sites.
Since $\mu$ is not known {\it a priori}, the relations for
$\mu > 0$, $\mu \in [-1, 0)$, [-2, -1), [-3, -2), [-4, -3), and $\mu < -4$
are all analyzed.   We solve all six zero-current conditions and determine which is self-consistent,
i.e., yields a $\mu$ value in its proper interval.
In practice, given the stationary values $\rho^\pm(j)$, we find that only one of the six conditions is in fact
satisfied in a self-consistent manner.

From the Ising-lattice gas correspondence we know that for $\rho = 1/2$, the chemical
potential $\mu = -2$ for all temperatures, if $D=0$.  The driven KLS lattice gas is invariant under
exchange of particles and empty sites or {\it holes} if we also let ${\bf D} \to -{\bf D}$.
Since the stationary properties depend on $D$, but not on the orientation of the drive,
particle-hole symmetry (PHS) also applies to the driven lattice gas.  In particular,
for $\rho = 1/2$, in the stationary state, we have the symmetries $\rho^-(j) = \rho^+(4-j)$,
for $j = 0,...,4$.
Inserting these relations in the zero-current condition, one finds that $\mu = -2$ for
density one half, in the stationary state, regardless of $T$ or $D$.
A further consequence of PHS is that, in both the driven and nondriven cases, if we define
$\Delta \mu(\rho,T,D) \equiv \mu(\rho,T,D) - \mu(1/2,T,D) = \mu(\rho,T,D) + 2$, then
$\Delta \mu$ is an odd function of $\Delta \rho \equiv \rho - 1/2$.

In the present study we restrict attention to temperatures above the critical temperature $T_c$.
For the driven lattice gas on the square lattice, $T_c \to 0.769(2)$ as
$D \to \infty$ \cite{marro87}, while the
corresponding critical temperature in the pair approximation (PA) is 0.8015 \cite{mftdds}.
We concentrate on a temperature of unity; qualitatively similar results are found at
other temperatures above $T_c$.

We study the KLS model on square lattices of $L \times L$ sites
with $L = 40$, again with periodic boundaries.  After verifying that the system has attained a steady state,
we determine the densities $\rho^\pm(j)$.
In a particle exchange move, a site $(i,j)_A$
in lattice A is chosen at random.
If the occupancy states of this site and the corresponding site $(i,j)_B$ are different,
then a particle is exchanged between the lattices with probability $p_{ex} = \min\{1, \exp(-\beta \Delta E)\}$
where the energy change $\Delta E$ involves nearest-neighbor particle interactions within each lattice,
but not the drive.  There are no interactions between particles in lattices A and B.
We use particle exchange probabilities $p_r = 0.0002$, 0.0005 and 0.001 and extrapolate to the
weak exchange limit.
Studies using {\it random exchange}, (i.e., between sites $(i,j)_A$ and $(i',j')_B$, chosen
independently in systems A and B), yield similar results, suggesting that stationary properties
are not sensitive to this choice.

We complement our simulation studies with PA analyses, following the
approach devised in \cite{mftdds}.  Briefly, the PA is formulated in terms of a set
of coupled (nonlinear) differential equations for the nearest-neighbor pair probabilities $b(i,o)$, where
$i$ denotes the pair type (doubly occupied, doubly vacant, or mixed), and $o$ denotes
the orientation (parallel or perpendicular to the drive).  Joint probabilities involving three or more
sites are approximated on the basis of the pair probabilities, as detailed in \cite{mftdds}.
We study homogeneous systems
at fixed temperature, density, and drive, integrating the equations until the stationary state
is attained.  Using the stationary pair probabilities we calculate the densities
$\rho^{\pm}(j)$ and hence the quantity $\mu^*$ via the zero-current condition.
We extend this method to study a pair of systems under weak exchange.

\begin{figure}[!htb]
\includegraphics[clip,angle=0,width=0.8\hsize]{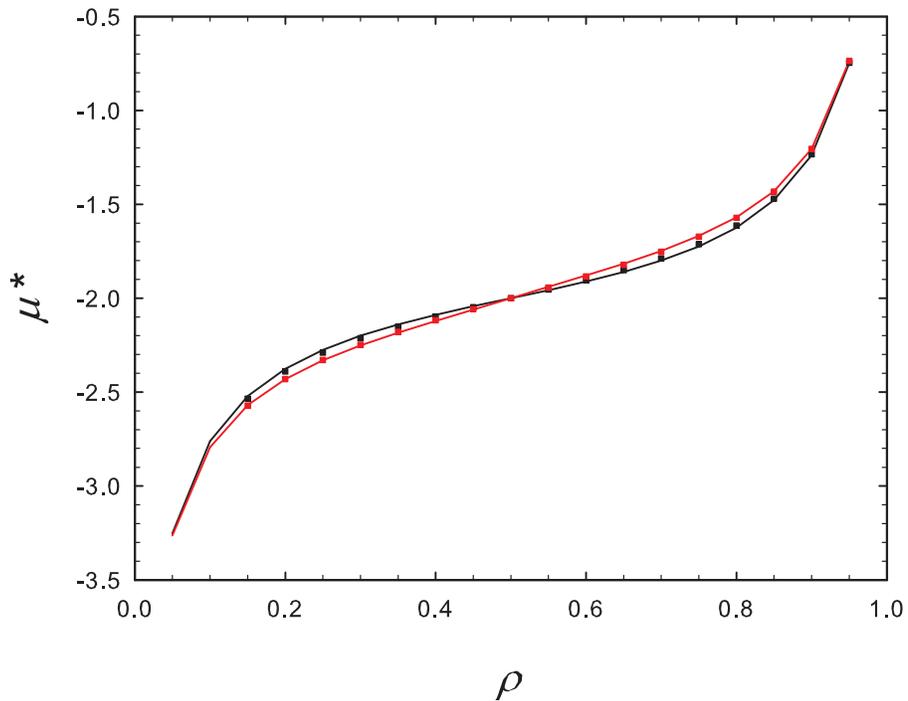}

\caption{(Color online) KLS lattice gas: simulation results (points, $L=40$), and PA predictions (curves) for
for $\mu^*$ in equilibrium (black)
and under a strong drive, $D=10$, (red), for temperature $T=1$.}
\label{muvrhT1}
\end{figure}

Figure~\ref{muvrhT1} shows simulation and PA results for $\mu^* (\rho,T,D)$
for the KLS model
in equilibrium and under a strong drive ($D=10$), for temperature $T=1$.  For $\rho < 1/2$,
$\mu^*$ is smaller in the driven system than in equilibrium, for the same density; for $\rho > 1/2$
the trends reverse.  The curves cross at $\rho = 1/2$, as expected.

\begin{figure}[!htb]
\includegraphics[clip,angle=0,width=0.8\hsize]{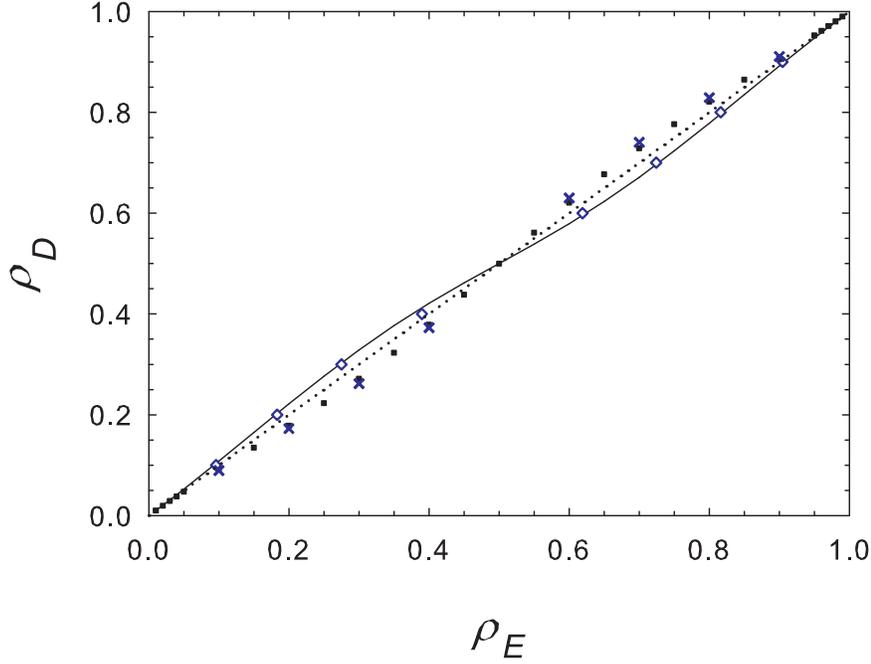}
\caption{\footnotesize{\sf (Color online) KLS lattice gas: PA predictions and simulation (MC) results
for coexisting densities $\rho_E$ (for $D\!=\!0$) and
$\rho_D$ (for $D\!=\!10$).  Equating chemical potentials in the driven and nondriven systems
yields the smooth curve (PA), and the blue diamonds (MC).
For systems under weak exchange, the corresponding values are denoted by black squares (PA)
and blue crosses (MC).
The dotted diagonal line corresponds to $\rho_D = \rho_E$.}}
\label{cmp34a}
\end{figure}

With the values of $\mu^*$ in hand, we study pairs of systems, one
in equilibrium, the other with $D=10$, under weak exchange.  The results shown in Fig.~\ref{muvrhT1}
lead one to expect that if the two systems are initialized with the same density,
$\overline{\rho} > 1/2$, then particles will
migrate from the driven to the nondriven system, until the corresponding $\mu^*$ values are the same.
In fact, particles flow in the {\it opposite} sense, so  that, in the stationary state,
$\rho (D\!=\!10) > \rho (D\!=\!0)$, and $\mu^* (D\!=\!10) > \mu^* (D\!=\!0)$.
This is demonstrated in Fig.~\ref{cmp34a}:
the curve shows the relation between coexisting densities $\rho_E$ and $\rho_D$ in the
equilibrium and driven systems, respectively, predicted by equating
$\mu^*(\rho,T,D\!=\!0)$ and $\mu^*(\rho,T,D\!=\!10)$, using the PA.
The small squares represent the coexisting densities given by the PA under weak exchange
($p_r = 0.001$).  The latter lie on the opposite side of the line $\rho_D = \rho_E$ from the curves
obtained using equal chemical potentials.  The simulation data (diamonds and crosses) follow the
same trends as the PA predictions.
We conclude that, using Metropolis exchange rates, equating values of $\mu^*$
does not predict coexisting densities in the KLS model.

\subsection{KLS lattice gas under Sasa-Tasaki contact}

We now consider the KLS model in contact with a reservoir or another KLS system using ST rates for
particle exchange.
Following the ST prescription, the dimensionless chemical potential of a system
(driven or not) of $N = \rho V$ particles on a lattice of $V$ sites is given by:

\begin{equation}
\mu^* = \ln \frac{g}{1-\rho}
\label{mustarST}
\end{equation}

\noindent where $g = \sum_{j=0}^q  \rho^-(j) e^{-\beta j}$ (see Appendix B of \cite{sasa2006}).
Sasa and Tasaki show that a pair of systems coexist under
exchange of particles when their respective values of $\mu^*$ are equal.

We verify the ST coexistence criterion in simulations
of the KLS model on a lattice with $L=40$ and temperature $T=1$.
To begin we determine $\mu^*$ in the nondriven system for densities near 0.75;
in a system with drive $D=10$, similar values of $\mu^*$ are found for
$\rho \simeq 0.606$.  Having obtained $\mu^*(\rho)$ for the two systems, we find that,
for a fixed total particle number of 2169, the dimensionless chemical potential
takes the common value of $\mu^* = -1.71315$ for densities 0.75004 and 0.60558 in the nondriven and
driven systems, respectively.  These are the predicted coexistence
densities for a pair of systems, one nondriven
and the other with $D=10$, each at temperature $T=1$.
Simulating the two systems under weak exchange, we verify that the stationary densities in fact
approach the predicted values as $p_r \to 0$, and that the chemical potentials
defined via Eq.~(\ref{mustarST}) approach the common value cited above.

Summarizing, we have shown that in the case of the KLS model, coexistence
under particle exchange is not predicted correctly using the condition of
equal dimensionless chemical potentials, if the exchange is governed by
Metropolis rates.  Correct predictions are however obtained using ST exchange rates.
We expect the same conclusions to hold for other interacting (nonathermal)
lattice gases.  We may further expect that other exchange rates (such as mechanism-B rates)
which do not share the ST property, will yield incorrect predictions.

\section{KLS lattice gas: effective temperature}

The results of Sec. IV demonstrate that equating the dimensionless chemical potentials
$\mu^*$, of a driven and an nondriven KLS system, does not predict the
stationary densities when these systems are allowed to exchange particles using Metropolis rates.
The reason for this becomes apparent when we examine the energy transfer
$\langle \Delta E\rangle_{{\cal S}{\cal S}_0}$
from the nondriven to the driven system at coexistence, under weak exchange.
A steady state requires $\langle \Delta n \rangle = 0$, but
the particle exchanges on average transfer energy from the driven to the
nondriven system, making $\langle \Delta E \rangle_{{\cal S}{\cal S}_0} < 0$.
(For the parameters studied here, $|\langle \Delta E \rangle_{{\cal S}{\cal S}_0}|$
is on the order of $10^{-3}$ to $10^{-2}$.)
A steady flux of energy from the driven to the nondriven system is also observed
under ST exchange rates, even though equating the values of $\mu^*$ yields
the coexisting densities correctly: the systems coexist under particle exchange but not under
energy exchange.
In this section we examine the possibility of predicting coexistence (under both particle and energy exchange)
using two parameters, by introducing
an effective temperature as well as an effective chemical potential.

The steady flow of energy from the driven to the nondriven system suggests that
the former is effectively at a higher temperature than the latter.
Note that in the driven KLS system ${\cal S}$, the acceptance probability
$p_a = \min\{1, \exp[-\beta(\Delta E - {\bf D} \cdot \Delta {\bf x})]\}$
implies contact with a reservoir at temperature $T = 1/\beta$;
since this reservoir is only accessible to ${\cal S}$, we call it the {\it private reservoir}, ${\cal R}_{\cal S}$.
On average, the drive increases the energy of ${\cal S}$, since it
increases the likelihood of transitions with $\Delta E > 0$ more than those with $\Delta E < 0$.
Under steady conditions, the energy increase due to the drive
is balanced by the energy transfer to ${\cal R}_{\cal S}$: this is the
{\it housekeeping heat} associated with the stationary
operation of a driven system \cite{oono-paniconi}.
Under the drive, we may regard the temperature of ${\cal R}_{\cal S}$
as {\it merely a parameter}
in the definition of the transition probabilities; we refer to it as the ``nominal temperature".
We expect the effective temperature, $T_e$, of the driven system
to be greater than $T_n$.

The parameters $\mu$ and $T_e$ of the KLS system will again be defined via exchange with a reservoir.
If a system ${\cal S}$ is in contact with a (heat and particle)
reservoir at temperature $T_e$ and chemical potential $\mu$, then, assuming Metropolis exchange rates,
an exchange in which the
energy of ${\cal S}$ changes by $\Delta E$, and the number of particles in ${\cal S}$ changes
by $\Delta n$, is accepted with probability $p = \min \{1, \exp[\beta_e(\mu \Delta n - \Delta E)] \}$,
where $\beta_e = 1/T_e$.  The effective temperature and chemical potential of ${\cal S}$ are
then determined by the conditions:

\begin{equation}
\langle \Delta n \rangle_{\cal S} = \sum_{j=0}^q \left[ \rho_{\cal S}^+(-j) \min\{1,e^{\beta_e(\mu + j)}\} -
                             \rho_{\cal S}^-(j) \min\{1,e^{-\beta_e(\mu + j)}\} \right] = 0
\label{delnzero1}
\end{equation}
and
\begin{equation}
\langle \Delta E \rangle_{\cal S} = -\sum_{j=0}^q j \left[ \rho_{\cal S}^+(-j) \min\{1,e^{\beta_e(\mu + j)}\} -
                             \rho_{\cal S}^-(j) \min\{1,e^{-\beta_e(\mu + j)}\} \right] = 0 .
\label{delezero1}
\end{equation}
We take $\rho_{\cal S}^+(-j)$ and $\rho_{\cal S}^-(j)$ as averages
over the stationary state in which ${\cal S}$ exchanges energy and particles with the reservoir;
in equilibrium, these densities therefore represent grand canonical averages.  If ${\cal S}$ is driven,
its stationary properties depend on the rate $p_r$ of exchange attempts with the reservoir, and
$\mu$ and $T_e$ are defined in the weak-exchange limit.

Our definitions of $\mu$ and $T_e$ for a driven system ${\cal S}$ are a direct consequence of the principle that,
if ${\cal S}$ coexist with a reservoir ${\cal R}$, it must be characterized by the same temperature and
chemical potential as ${\cal R}$.  We now ask whether this definition is thermodynamically consistent.
Given a reservoir ${\cal R}$ such that Eqs.~(\ref{delnzero1})
and (\ref{delezero1}) hold for some driven system ${\cal S}$, let ${\cal S}_0$ be an equilibrium
KLS lattice gas with the same temperature and chemical potential as ${\cal R}$.
Since ${\cal S}_0$ and ${\cal R}$ are
in equilibrium, we have, for the site densities $\rho_0^{\pm} (j)$ of ${\cal S}_0$,
the detailed-balance relations,
\begin{equation}
\rho_0^+ (-j) = e^{-\beta_e (\mu + j)} \rho_0^-(j),
\label{detbal}
\end{equation}
\noindent
which of course imply $\langle \Delta n \rangle_0 = \langle \Delta E \rangle_0 = 0$ for exchange between
${\cal S}_0$ and ${\cal R}$.

Consistency requires that ${\cal S}$ and ${\cal S}_0$ coexist, that is, that the particle and energy fluxes
between these systems be zero under virtual exchange.
These fluxes (per exchange attempt, from the nondriven to the driven system)
are given by,
\begin{equation}
\langle \Delta n \rangle_{{\cal S}{\cal S}_0} = \sum_{j,j_0} \left[ \rho_{\cal S}^+(-j) \rho_0^-(j_0)
\min\{1,e^{\beta_e(j-j_0)}\} -
                             \rho_{\cal S}^-(j) \rho_0^+ (-j_0) \min\{1,e^{\beta_e(j_0 - j)}\} \right]
\label{delnss0}
\end{equation}
and
\begin{equation}
\langle \Delta E \rangle_{{\cal S}{\cal S}_0} = -\sum_{j,j_0} j
\left[ \rho_{\cal S}^+(-j) \rho_0^- (j_0) \min\{1,e^{\beta_e(j - j_0)}\} -
                             \rho_{\cal S}^-(j) \rho_0^+ (-j_0) \min\{1,e^{\beta_e(j_0 - j)}\} \right].
\label{deless0}
\end{equation}

\noindent Using detailed balance we may write,
\begin{eqnarray}
\langle \Delta n \rangle_{{\cal S}{\cal S}_0} \!\!\!\!\!\! &=& \!\!\!\!\!\!\!\! \sum_{j_0} \left\{
\sum_{j<j_0} \left[ \rho_{\cal S}^+(-j) \rho_0^-(j_0) e^{\beta_e(j-j_0)} -
\rho_{\cal S}^-(j) \rho_0^+ (-j_0) \right] \right.
\nonumber
\\
&\;\;\;\;\;\;\;\;\;\;\;\;+& \left.
\sum_{j \geq j_0} \left[ \rho_{\cal S}^+(-j) \rho_0^-(j_0)  -
\rho_{\cal S}^-(j) \rho_0^+ (-j_0) e^{\beta_e(j_0-j)} \right] \right\}
\nonumber
\\
&=& \!\!\!\!\!\!\!\! \sum_{j_0} \rho_0^- (j_0) \left\{
\sum_{j<j_0} e^{-\beta_e(j_0-j)}
\left[ \rho_{\cal S}^+(-j)  -
e^{-\beta_e(\mu+j)} \rho_{\cal S}^-(j)  \right] \right.
\nonumber
\\
&\;\;\;\;\;\;\;\;\;\;\;\;+&
\left.
\sum_{j \geq j_0} \left[ \rho_{\cal S}^+(-j) -
 e^{-\beta_e(\mu+j)} \rho_{\cal S}^-(j) \right] \right\}
\nonumber
\\
&=& \!\!\!\!\!\!\!\! \sum_{j_0} \alpha(j)
\left[ \rho_{\cal S}^+(-j)  -
e^{-\beta_e(\mu+j)} \rho_{\cal S}^-(j)  \right]
\label{delnss01}
\end{eqnarray}
and
\begin{equation}
\langle \Delta E \rangle_{{\cal S}{\cal S}_0}
= -\sum_{j_0} j \, \alpha(j)
\left[ \rho_{\cal S}^+(-j)  -
e^{-\beta_e(\mu+j)} \rho_{\cal S}^-(j)  \right],
\label{deless01}
\end{equation}
where
\begin{equation}
\alpha(j) \equiv \sum_{k=0}^j \rho_0^-(k) + \sum_{k=j+1}^q e^{\beta_e (j-k)} \rho_0^-(k) .
\label{alphaj}
\end{equation}

Now $\alpha(j)$ is an increasing function of $j$, and since it takes a different value for each $j$, the conditions
$\langle \Delta n \rangle_{{\cal S}{\cal S}_0} = 0$ and
$\langle \Delta E \rangle_{{\cal S}{\cal S}_0} = 0$
are in general distinct from Eqs.~(\ref{delnzero1}) and (\ref{delezero1}), which define $\mu$ and $T_e$.
The four conditions cannot be
satisfied simultaneously except in special cases.  For example, for $\mu > 0$,
$\langle \Delta n \rangle_{{\cal S}} = 0$ implies that
$\sum_{j_0} \left[ \rho_{\cal S}^+(-j)  -
e^{-\beta_e(\mu+j)} \rho_{\cal S}^-(j)  \right] = 0$,
which is clearly different from the condition $\langle \Delta n \rangle_{{\cal S}{\cal S}_0} = 0$.
We have therefore demonstrated a violation of the zeroth law: although ${\cal S}$ and ${\cal S}_0$
both coexist with ${\cal R}$, in general they do not coexist with each other.
Evidently this conclusion holds for
attractive nearest-neighbor lattice gases on any regular lattice, in
dimension $d \geq 2$, regardless of the coordination number.

The argument can be extended to other functional forms of the acceptance
probability $p_a$.
For example, in the case of {\it mechanism B rates},
a transition involving a particle displacement $\Delta {\bf x}$,
attended by a change $\Delta E$ in interaction energy, is accepted with probability
\begin{equation}
p_B = \frac{1}{1+  \exp[\beta(\Delta E - {\bf D} \cdot \Delta {\bf x})]},
\label{padB}
\end{equation}
with $\beta = 1/T_n$,
while for particle and energy exchange with a reservoir at inverse temperature $\beta_e$ and
chemical potential $\mu$ we have
\begin{equation}
p_B = \frac{1}{1+  \exp[\beta_e(\Delta E - \mu \Delta n)]}.
\label{padBR}
\end{equation}

\noindent The parameters $\beta_e$ and $\mu$ characterizing the driven system ${\cal S}$ satisfy
the relations,
\begin{equation}
\langle \Delta n \rangle_{\cal S} = \sum_j \left[\frac{\rho_{\cal S}^+(-j)}{1+e^{-\beta_e(\mu+j)}}
- \frac{\rho_{\cal S}^-(j)}{1+e^{\beta_e(\mu+j)}} \right] = 0
\label{delnB}
\end{equation}
and
\begin{equation}
\langle \Delta E \rangle_{\cal S} = -\sum_j j \left[\frac{\rho_{\cal S}^+(-j)}{1+e^{-\beta_e(\mu+j)}}
- \frac{\rho_{\cal S}^-(j)}{1+e^{\beta_e(\mu+j)}} \right] = 0.
\label{delEB}
\end{equation}

\noindent Using the relation analogous to Eq. (\ref{delnss0}), with mechanism B acceptance probabilities in place
of the Metropolis expressions, and Eq. (\ref{detbal}), one may now write,

\begin{equation}
\langle \Delta n \rangle_{{\cal S}{\cal S}_0} = \sum_{j_0} \rho_0^-(j_0)
\sum_j \frac{1+e^{-\beta_e(\mu+j)}}{1+e^{\beta_e(j_0-j)}}
\left[\frac{\rho_{\cal S}^+(-j)}{1+e^{-\beta_e(\mu+j)}}
- \frac{\rho_{\cal S}^-(j)}{1+e^{\beta_e(\mu+j)}} \right] .
\label{delnss0B}
\end{equation}

\noindent Once again, it is not possible to satisfy all four coexistence conditions
simultaneously.

Here it is worth noting that the zero-current conditions are
qualitatively different for different choices of $p_a$, so that the effective temperature
and chemical potential of the driven system will depend on the choice of rates.
What is more, for a KLS lattice gas with $p_a$ given as above, $\mu$ and $T_e$ depend on whether
we use Metropolis, mechanism B, or some other acceptance probabilities for exchanges with the
reservoir.

As was shown in Sec. IV, coexistence under particle exchange using ST rates is
predicted correctly and consistently using equality of $\mu^*$.
There remains, nevertheless, a steady transfer of energy from the driven to the nondriven
system in this case.
(Numerical values of the flux again lie in the range $10^{-3}$ to $10^{-2}$.)
As before, we define an effective temperature for the driven
system via the zero-energy-flux condition.
For ST rates, the zero-flux conditions determining
the effective temperature and chemical potential of ${\cal S}$ become:

\begin{equation}
\langle \Delta n \rangle_{\cal S} = \sum_{j=0}^q \left[ \rho_{\cal S}^+(-j) -
                             \rho_{\cal S}^-(j) e^{-\beta_e(\mu + j)} \right] = 0
\label{delnzeroTS}
\end{equation}
and
\begin{equation}
\langle \Delta E \rangle_{\cal S} = -\sum_{j=0}^q j \left[ \rho_{\cal S}^+(-j) -
                             \rho_{\cal S}^-(j) e^{-\beta_e(\mu + j)} \right] = 0 .
\label{delezeroTS}
\end{equation}

\noindent Solving each equation for $e^{\mu^*}$
and equating the resulting expressions we obtain

\begin{equation}
\frac{\sum_j j \rho_{\cal S}^-(j) e^{-\beta_e j}}{\sum_j \rho_{\cal S}^-(j) e^{-\beta_e j}}
= \frac{\sum_j j \rho_{\cal S}^+(-j)}{1-\rho}
\label{teffST}
\end{equation}
\noindent which can be solved for the effective temperature, $T_{e} = 1/\beta_e$, of the driven system;
then Eq. (\ref{delnzeroTS}) defines its dimensionless chemical potential, $\mu^*$.

In this case, the inconsistency discussed above is absent.
If ${\cal S}$ and ${\cal S}_0$ are driven and nondriven systems coexisting with a reservoir
having parameters $\mu^*$ and $\beta_e$,
the fluxes between driven and nondriven systems are given by,

\begin{equation}
\langle \Delta n \rangle_{{\cal S}{\cal S}_0} = \sum_{j,j_0} \left[ \rho_{\cal S}^+(-j) \rho_0^-(j_0)
e^{-\beta_e j_0} -
                             \rho_{\cal S}^-(j) \rho_0^+ (-j_0) e^{-\beta_e j} \right]
\label{delnssTS}
\end{equation}
and
\begin{equation}
\langle \Delta E \rangle_{{\cal S}{\cal S}_0} = -\sum_{j,j_0} j
\left[ \rho_{\cal S}^+(-j) \rho_0^- (j_0) e^{-\beta_e j_0} -
                             \rho_{\cal S}^-(j) \rho_0^+ (-j_0) e^{-\beta_e j} \right].
\label{delessTS}
\end{equation}

\noindent Now using detailed balance we may write,

\begin{equation}
\langle \Delta n \rangle_{{\cal S}{\cal S}_0} = \sum_{j_0} e^{-\beta_e j_0} \rho_0^-(j_0)
\sum_j \left[ \rho_{\cal S}^+(-j) - e^{-\beta_e (\mu+j)} \rho_{\cal S}^-(j) \right],
\label{delnssTS1}
\end{equation}

\noindent so that $\langle \Delta n \rangle_{\cal S} = 0$ implies $\langle \Delta n \rangle_{{\cal S}{\cal S}_0} = 0$,
and similarly, $\langle \Delta E \rangle_{\cal S} = 0$ implies $\langle \Delta E \rangle_{{\cal S}{\cal S}_0} = 0$.
Thus rates enjoying the ST property do not suffer from the inconsistency discussed above.

\subsection{Numerical examples}

\begin{figure}[!htb]
\includegraphics[clip,angle=0,width=0.6\hsize]{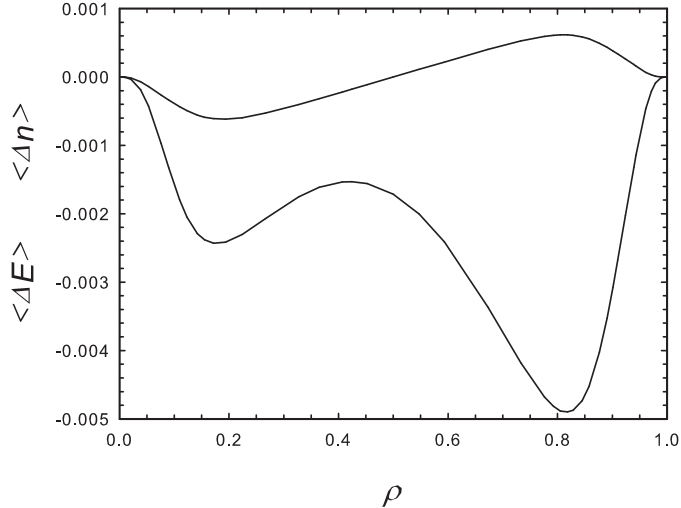}

\caption{\footnotesize{\sf KLS lattice gas: particle flux $\langle \Delta n \rangle_{{\cal S}{\cal S}_0}$ (upper curve)
and energy flux $\langle \Delta E \rangle_{{\cal S}{\cal S}_0}$ (lower curve), under virtual exchange, between
a driven system ${\cal S}$ (nominal temperature $T_n=0.6$, drive $D=10$) and an equilibrium system ${\cal S}_0$, both of which coexist with the
same reservoir.  Results obtained via numerical solution of the master equation for systems of 4$\times$4 sites.
}}
\label{dlne6}
\end{figure}

Although we have shown analytically that an effective temperature and chemical potential
cannot be defined in a consistent manner for the KLS model under Metropolis exchange rates,
it is interesting to study numerical examples, to assess the degree
of inconsistency.
We obtain the stationary solution to the master equation for the KLS model,
including exchange with a reservoir at temperature $T_e$ and chemical potential $\mu$, on a lattice of
$4 \times 4$ sites, using the method of \cite{qsexact}. In the equilibrium case we verify that
$\langle \Delta E \rangle =0$ for $T_e = T_n$, independent of the exchange rate $p_r$, as expected.
For a nonzero drive, we determine $T_e$ such that $\langle \Delta E \rangle =0$
for a series of $p_r$ values, permitting extrapolation to $p_r =0$.
(For the range of values used, $p_r = 0.001 - 0.006$,
all quantities of interest are essentially linear functions of $p_r$.)
With the values of $\mu$, $T_e$ and
the $\rho_{\cal S}^+(-j)$ and $\rho_{\cal S}^-(j)$ in hand, we determine the corresponding densities in an
equilibrium KLS system at temperature $T_e$ and chemical potential $\mu$.  This permits us to calculate
the fluxes under virtual exchange using Eqs.~(\ref{delnss0}) and (\ref{deless0}).
For $0.6 \leq T_n \leq 1$, we observe significant fluxes (of order $10^{-3} - 10^{-4}$)
between the driven system (with $D=10$) and the corresponding nondriven one.
The density dependence of the particle and energy fluxes for $T_n =0.6$ and $D=10$ is shown
in Fig.~\ref{dlne6}.
For the parameters of this study, $|\langle \Delta E \rangle_{{\cal S}{\cal S}_0}|$ takes its
maximum value (slightly less than $0.005$) for $\rho \simeq 0.805$, which corresponds to $T_e \simeq 0.845$;
the energy flux is equivalent to an {\it equilibrium} temperature mismatch of about 0.4\% in the
KLS model with these parameters.  Studies using $T_n=1$ yield somewhat smaller fluxes, i.e.,
$|\langle \Delta E \rangle_{{\cal S}{\cal S}_0}| \sim 5 \times 10^{-4}$ and
$|\langle \Delta n \rangle_{{\cal S}{\cal S}_0}| \sim 10^{-4}$.

Simulations of driven and nondriven KLS systems in contact, following the same scheme as in
the analysis of the master equation, confirm the presence of inconsistencies.  For system sizes $L=20$ and 40,
and $T_n=1$, the associated fluxes are similar to those found for $L=4$.
In summary, systematic violations
of consistency are observed in studies using Metropolis exchange rates in determining the effective parameters
and for exchange between driven and nondriven systems.

By contrast, simulations using ST exchange rates verify
the consistency of the effective temperature and chemical potential
of driven systems.  In these studies, we first simulate the driven system (with fixed particle number,
$N_{\cal S}$) to determine
the $\rho_{\cal S}^\pm(\pm j)$, and obtain $T_e$ and $\mu^*$ via Eqs. (\ref{teffST}) and (\ref{delnzeroTS}).
Next we simulate a nondriven system at temperature $T_e$, for several values
of the particle number $N_0$, to determine the value for which $\mu_0^*$ is equal to $\mu^*$ of the
driven system.
For these parameters, ${\cal S}$ and ${\cal S}_0$ evidently coexist with the same reservoir.  Finally, to test for
consistency, we simulate the two systems under weak exchange.  We verify that in the
steady state, the particle numbers are those obtained via the coexistence criteria, and that the stationary
energy flux between the systems is zero to within uncertainty.

A study using $L=320$, $T_n=1$, $D=10$ and $\rho_{\cal S} = 0.6$ (particle number $N_{\cal S} = 61440$)
yields $T_e = 1.72527(5)$ and $\mu^* = -0.954049(6)$ for the driven system;
the corresponding nondriven system (at temperature $T_e$) has density
$\rho_0 = 0.60043$ (particle number $N_0 = 61484$) and $\mu^* = -0.954050(1)$, equal to within uncertainty
to that of the driven system.  Studies of the two systems under weak exchange ($p_r = 0.0005$, 0.001 and 0.002)
yield the limiting values $\mu_0^* = -0.9541(2)$ and $\mu_{\cal S}^* = -0.9539(2)$, and densities
$\rho_0 = 0.60042(5)$ and $\rho_{\cal S} = 0.60000(5)$.  Thus the properties of the coexisting systems
agree with those predicted by equating the effective temperature
and dimensionless chemical potential of the driven and nondriven systems in isolation.
In the weak exchange limit the energy flux
$\langle \Delta E \rangle_{{\cal S}{\cal S}_0}$ is zero to a precision of $10^{-5}$.  Similar confirmations
of consistency are found for systems with $L=80$ and 160.

\begin{figure}[!htb]
\includegraphics[clip,angle=0,width=0.8\hsize]{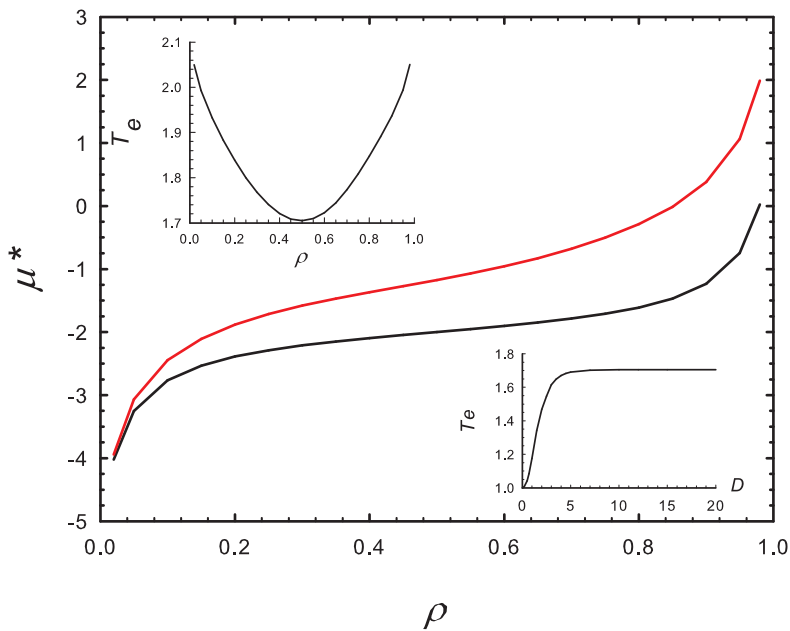}

\caption{\footnotesize{\sf KLS lattice gas: dimensionless chemical potential $\mu^*$
versus density for $T=1$ in equilibrium
(black curve) and effective $\mu^*$ for driven lattice gas with $T_n=1$
and drive $D=10$.  Upper inset: effective temperature $T_e$ of driven system versus density.  Lower inset:
effective temperature of driven system versus drive $D$ for density $\rho =0.5$ and $T_n=1$.  System size
$L=80$.  Effective parameters for the driven system are obtained using Sasa-Tasaki exchange rates.
}}
\label{muteST}
\end{figure}

It is of interest to examine how the effective parameters of the driven system, obtained under ST exchange
with the reservoir, depend upon density and drive strength; some examples are shown in Fig.~\ref{muteST}.
The dimensionless chemical potential of the driven system follows the same general trend, as a function of
density, as in equilibrium, but is consistently greater; the difference is an increasing function of density.
The effective temperature is always greater than $T_n$, as expected.  It exhibits a minimum at half-filling.
For fixed density, the excess effective temperature $T_e - T_n$ exhibits a sigmoidal form, growing
$\propto D^2$ for a weak drive, then roughly linearly, before saturating for $D \approx 10$.

\section{Conclusions}

We have examined, in concrete, operational terms, the consistency of steady state
thermodynamics (SST) for driven lattice gases.  Effective
intensive parameters (temperature and chemical potential) for the driven system
are defined via zero-flux conditions under
weak exchange with a reservoir.  Consistency requires that the parameters obey the zeroth law.

In the case of the lattice gas with nearest-neighbor exclusion, only particle
fluxes are of interest.  The effective chemical potential is consistent with
the zeroth law, and correctly predicts
the coexisting densities of systems with distinct values of the drive,
as verified for the NNE lattice gas with nearest-neighbor hopping only, and in a
second model which includes hopping to second neighbors as well.  We expect that this
will be true of other models with purely excluded-volume interactions, such as lattice gases
with extended hard cores or the hard-sphere fluid.

Analysis of the driven lattice gas with nearest-neighbor
attractive interactions (the KLS model) shows that we must define an effective
temperature $T_e$ as well as an effective chemical potential.
We then explore the possibility of predicting coexistence
using equality of these parameters.
A theoretical argument shows that the zeroth law is violated for generic exchange rates,
but {\it not} for Sasa-Tasaki rates.
Exact solutions of small systems, and simulations of larger systems,
reveal that above the critical temperature, the discrepancies obtained using Metropolis rates,
while significant, are not very large, consistent with the findings of \cite{pradhan}.
The possibility of nonzero particle and energy fluxes between the driven (${\cal S}$)
and nondriven (${\cal S}_0$) systems,
even when they coexist with the {\it same reservoir}, arises because energy flows continuously
from the drive, through ${\cal S}$ and thence to
its private reservoir ${\cal R}_{\cal S}$.
Simulations confirm the zeroth law for Sasa-Tasaki rates.


Our results suggest that
extending thermodynamics, fully and consistently, to
nonequilibrium steady states requires using rates with the ST property:
the rate for particle transfer from system $A$ to system $B$ depends exclusively on parameters
associated with $A$, and vice-versa.  Physically, ST rates correspond to
the initial and final particle positions being separated by a high energy barrier.
While this seems eminently reasonable for many systems of potential interest, it is not yet clear
if this principle applies universally.

Note that in the KLS system we verify consistency using ST rates
for exchanges between systems (and between system and reservoir), {\it even though the internal dynamics
of each system follows Metropolis rates}.  We believe that for single-phase systems, the ST exchange scheme
will yield consistent definitions of the effective intensive parameters
for {\it any} choice of the rates governing the internal dynamics,
as long as local detailed balance is satisfied.  Naturally, one hopes to apply SST to study phase coexistence
within a {\it single} system; here it seems likely that a consistent description will require ST rates
for the internal dynamics as well.  We intend to address this issue in the near future.

A number of other questions remain for future investigation.  Applications to
athermal models in continuous space, and to models including particle momentum variables,
are of interest in extending the analysis to more realistic systems.  A further question is
whether an entropy function can be constructed for NESS.

\vspace{2em}

\noindent {\bf Acknowledgments}

We thank J\"urgen Stilck for helpful comments.
This work was supported by CNPq and CAPES, Brazil.

\end{document}